\documentclass[
 reprint,
 amsmath,amssymb,
 aps,
prb,
]{revtex4-2}

\usepackage{graphicx}% Include figure files
\usepackage{dcolumn}% Align table columns on decimal point
\usepackage{bm}% bold math
\usepackage{hyperref}% add hypertext capabilities
\usepackage{tikz}
\usepackage{comment}

\DeclareMathOperator{\Tr}{Tr}
\let\Re\relax
\let\Im\relax
\DeclareMathOperator{\Re}{Re}
\DeclareMathOperator{\Im}{Im}
\DeclareMathOperator{\dd}{d}
\newcommand{\matr}[1]{\underline{\underline{#1}}}

\begin{document}

\preprint{APS/123-QED}

\title{Reconciling global and local spin models}

\author{Bendegúz Nyári$^{1,2}$}
%\email{nyari.bendeguz@ttk.bme.hu}
\author{András Lászlóffy$^{3,4}$}
\author{László Szunyogh$^{1,2}$}
%\email{szunyogh.laszlo@ttk.bme.hu}
\affiliation{$^1$Department of Theoretical Physics, Institute of Physics, Budapest University of Technology and Economics, M\H{u}egyetem rkp.~3., HU-1111 Budapest, Hungary}
\affiliation{$^2$HUN-REN-BME Condensed Matter Research Group, Budapest University of Technology and Economics, M\H{u}egyetem rkp.~3., HU-1111 Budapest, Hungary}
\affiliation{$^3$Faculty of Information Technology and Bionics, Pázmány Péter Catholic University, 1083 Práter u. 50/A, Budapest, Hungary}
\affiliation{$^4$HUN-REN Wigner Research Centre for Physics, Institute for Solid State Physics and Optics, H-1525 Budapest, Hungary}

\date{\today}
\begin{abstract}
Magnetic interactions based on first-principles calculations can be casted either into global spin models, where the interactions are configuration-independent, or into local spin models, where the effective parameters depend explicitly on the underlying spin configuration. Although these two perspectives often yield seemingly inconsistent results, we demonstrate that they can be reconciled by restricting the analysis to transverse spin fluctuations. We develop a projection scheme that systematically removes spurious longitudinal contributions from local interaction matrices, thereby establishing a rigorous connection between the projected local interactions and the underlying global multispin Hamiltonian. Applying our approach to the prototype systems (i) a Cr trimer on Au(111) (ii) Fe and Cr dimers on Cu(001), we show that the projected local interaction parameters quantitatively reproduce those obtained from the global four-spin model. This work provides a unified framework to interpret local and global spin models on equal footing and resolves long-standing ambiguities in the identification of effective interactions such as the Dzyaloshinskii-Moriya interaction (DMI) in a non-collinear framework. %Our method offers a robust route to extract reliable effective spin Hamiltonians from ab initio calculations for complex non-collinear magnets.
\end{abstract}

%Use showkeys class option if keyword display desired
\maketitle

\section{Introduction}

Effective spin models provide a powerful framework for bridging first-principles electronic-structure calculations with the collective phenomena in magnetic materials. Since the early formulation of the Heisenberg Hamiltonian~\cite{Heisenberg1928}, such models have been indispensable in describing magnetic ordering, excitation spectra and finite temperature magnetism in solid magnets, for a comprehensive review see Ref.~\cite{SzilvaRevModPhys2023}. Evaluation of exchange interactions using Green function-based methods~\cite{Liechtenstein1987local} has played a crucial role in the advancement of the study of bulk magnets and magnetic systems in reduced dimensions, including thin films and interfaces~\cite{udvardi2003first}, as well as complex magnetic textures such as domain walls~\cite{Thiaville2002,Beach2005} and magnetic skyrmions~\cite{heinze2011spontaneous,Fert2017,tokura2020magnetic}. 

To capture the richness of competing interactions in these systems, extensions of the bilinear Heisenberg model have been proposed. In particular, multispin Hamiltonians including four-spin interactions~\cite{mankovsky2020extension,brinker2019chiral,gutzeit2022nano} provide a configuration-independent description that naturally accounts for the stabilization of complex non-collinear states. Such global spin models have been successfully applied to a wide range of magnetic systems, from chiral magnets to atomic clusters at the nanoscale \cite{heinze2011spontaneous,hayami2017effective,romming2018competition,kronlein2018magnetic,brinker2019chiral,paul2020role,khanh2020nanometric,tokura2020magnetic,hoffmann2020systematic,grytsiuk2020topological,brinker2020prospecting,gutzeit2022nano}.  

In the past decade, local spin models based on infinitesimal rotations~\cite{Szilva2013,Kvashnin2016,Rodrigues216} were developed, in which effective interactions explicitly depend on the underlying spin configuration. Although these models have provided useful insight into complex non-collinear states, they also led to {\em surprising} results, such as the emergence of Dzyaloshinskii–Moriya–like terms in the absence of spin–orbit coupling~\cite{cardias2020dzyaloshinskii,cardias2020first}. This has sparked an active debate regarding the consistency between local and global formulations and the possibility of unifying them within a coherent theoretical framework.~\cite{dos2021proper,cardias2022comment,dos2022reply,LaszloffyPhd}.  

In this work, we address this issue by establishing a direct connection between global and local spin models. Within a non-relativistic multiple scattering framework, in Section II we outline a method that enables the evaluation of two-spin and four-spin interactions in a global spin model and then in Section III we describe the ab initio method to calculate local exchange interactions in terms of infinitesimal rotations, whereby we introduce two formalisms providing identical two-spin variation of the energy. Applying these schemes to the prototypical case of a Cr trimer on Au(111), in Section IV we show that the two alternative formalisms yield remarkably different local magnetic interactions. In Section V, we present local interactions based on the previously introduced two-spin and four-spin global interactions and demonstrate that the local interactions are not uniquely defined due to longitudinal terms in the two-spin variation of the energy. In Section VI we use a projection method that systematically removes spurious longitudinal contributions from local interaction matrices and give evidence that the projected local interactions are well-defined, i.e., do not depend on the local formalism introduced before. Moreover, the projected local interactions are reproduced by the global four-spin Hamiltonian with high accuracy. In Section VII we extend the study for Fe and Cr dimers on Cu(001), then in Section VIII we summarize by concluding that local magnetic interactions straightforwardly descend from global multi-spin interactions, thereby giving no additional insight to the physics of magnetic systems.   

%Our results provide a unified and physically transparent framework that resolves ambiguities in the interpretation of effective spin interactions derived from first-principles.   

\section{Isotropic global spin model from first principles}
\label{sec:glob4spin}
The most general isotropic spin Hamiltonian containing two-spin and four-spin interactions can be written as
%\begin{widetext}
\begin{align}
    & H =\dfrac{1}{2}\sum_{ij}I_{ij}\vec{e}_i\vec{e}_j \nonumber \\
    & +\dfrac{1}{4}\sum_{ijkl}I_{ijkl}\left[ 
    (\vec{e}_i\vec{e}_j)(\vec{e}_k\vec{e}_l)-
    (\vec{e}_i\vec{e}_l)(\vec{e}_j\vec{e}_k)+
    (\vec{e}_i\vec{e}_k)(\vec{e}_j\vec{e}_l)
    \right] \, ,
    \label{eq:4spin_iso_ham}
\end{align}
%\end{widetext}
where the parameters $I_{ij}$ and $I_{ijkl}$ can be evaluated using multiple scattering theory based on the expansion of the grand potential in the spin variables up to the fourth order. The grand potential is given by Lloyd's formula \cite{lloyd1967wave,udvardi2003first,Lounis2020},
\begin{equation}
    \Omega(\{ \vec{e}\})=\Omega_0-\dfrac{1}{\pi}\Im\int_{-\infty}^{E_F} \dd \varepsilon \Tr \ln \boldsymbol{\tau}(\varepsilon,\{\vec{e}\})\;,
    \label{eq:lloyd}
\end{equation}
where $\Omega_0$ is the grand potential of the free electrons, $E_{\rm F}$ is the Fermi energy, and boldface quantities denote matrices in  site, angular momentum and spin indices. The scattering path operator (SPO) matrix $\boldsymbol{\tau}(\varepsilon,\{\vec{e}\})$ is defined as
\begin{equation}
    \boldsymbol{\tau}(\varepsilon;\{\vec{e}\})=(\mathbf{t}(\varepsilon;\{\vec{e}\})^{-1}-\mathbf{G}_0(\varepsilon))^{-1} \, 
    \label{eq:lntau}
\end{equation}
where $\mathbf{t}(\varepsilon;\{\vec{e}\})$ collects the single-site scattering matrices and $\mathbf{G}_0(\varepsilon)$ is the matrix of the free space structure constants. 
It should be mentioned that we considered only the single-particle energy of the electron system. This approach is based on the magnetic local force theorem which is strictly valid only to first order in the change of the total energy at a stationary state.\cite{Liechtenstein1987local} Importantly, the derivation of the global spin-model parameters is based on the same theoretical foundations as the torque method presented in the following section for determining the parameters of a local spin model.
%that can be expanded with respect to any reference system defined by the single-site scattering matrices ${\bf t}_r$. 

To obtain a systematic expansion in terms of spin variables, we decompose the $t$-matrix at site $i$ in angular-momentum and spin-space as
\begin{equation}
    \underline{\underline{t}}^i=\underline{t}^i_{r} I_2 + \underline{t}_s^i \vec{e}_i \vec{\sigma}\;,
    \label{eq:tspin}
\end{equation}
where in the first term $\underline{t}^i_{r}$ is the $t$-matrix of the non-magnetic reference system, $I_2$ is the two-by-two unit matrix and the second term stands for the perturbation due to spin polarization, with $\vec{\sigma}$ being the vector of the Pauli matrices and %$\vec{e}_i$ the direction of the magnetization at site $i$ 
we dropped the energy argument of the corresponding matrices. Note that Eq.~\eqref{eq:tspin} is valid only within a non-relativistic theory and the single underline refers to matrices with angular momentum indices $(\ell,m)$ only, while double underlines denote matrices with spin and angular momentum indices.

Defining the so-called structural Green's function matrix of the non-magnetic reference system,
\begin{equation}
    \mathbf{G}_r = \mathbf{G}_0(\mathbf{I}-\mathbf{t}_r\mathbf{G}_0)^{-1}\:,
    \label{eq:Gr}
\end{equation}
we can express the logarithm of the SPO matrix as
\begin{equation}
    \ln \boldsymbol{\tau} = \ln \mathbf{t} + \ln(\mathbf{I}-\mathbf{t}_r \mathbf{G}_0) +  \ln(\mathbf{I}-\mathbf{t}_s \mathbf{G}_r) \:.
    \label{eq:ln-split}
\end{equation}
Performing the trace in spin space %in Eq.~\eqref{eq:lloyd} From 
in the third term of the above equation, 
\begin{equation}
   \ln(\mathbf{I}-\mathbf{tG}_r)=-\sum_{k=1}^\infty\dfrac{1}{k} (\mathbf{t}_s \mathbf{G}_r)^k \, ,
   \label{eq:ln1mx}
\end{equation}
while keeping only terms up to $k=4$ only, we end up with Eq.~\eqref{eq:4spin_iso_ham}  (see Appendix \ref{appendix1}), where the two-spin and four-spin exchange interactions are defined as
\begin{align}
    I_{ij}&=-\dfrac{2}{\pi}\Im\int_{-\infty}^{E_F} \dd \varepsilon \Tr_L(\underline{t}^i_s \underline{G}_r^{ij}\underline{t}^j_s \underline{G}_r^{ji}) \label{eq:Iij} \\
    I_{ijkl}&=-\dfrac{2}{\pi}\Im\int_{-\infty}^{E_F} \dd \varepsilon \Tr_L(\underline{t}^i_s \underline{G}_r^{ij}\underline{t}^j_s \underline{G}_r^{jk}\underline{t}^k_s \underline{G}_r^{kl}\underline{t}^l_s \underline{G}_r^{li})\;. \label{eq:Iijkl}
\end{align}
%This way we obtained an \textit{ab initio} method that provides with two-spin and four-spin interactions.

It should be emphasized that the sums in Eq.~\eqref{eq:4spin_iso_ham} include all site indices. Using such a spin model in Monte Carlo or spin dynamics simulations demands to rearrange the Hamiltonian in terms of multisite interactions to avoid the coincidence of the sites in the summations. One can do this by systematically studying all possible index contractions in the Hamiltonian.  %The details of this rearrangement is discussed in Appendix \ref{appendix2}. 
The multisite Hamiltonian %including two-spin and four-spin interactions 
then reads as
%\begin{widetext}
\begin{align}
    H&=-\dfrac{1}{2} \sum_{ij}\! ' %\sum_{i\neq j}
    \: J_{ij}(\vec{e}_i\vec{e}_j)
    +\dfrac{1}{4} \sum_{ij}\!' %\sum_{i\neq j} 
    \: B_{ij}(\vec{e}_i\vec{e}_j)^2 \nonumber \\
    &+\dfrac{1}{4} \sum_{ijk}\!'  %\sum_{i\neq j\neq k} 
    \: T_{ijk} (\vec{e}_i\vec{e}_k)(\vec{e}_j\vec{e}_k) 
    +\dfrac{1}{4} \sum_{ijkl}\!'  %\sum_{i\neq j\neq k\neq l}
    \: K_{ijkl} (\vec{e}_i\vec{e}_j)(\vec{e}_k\vec{e}_l) \, ,
    \label{eq:4spin_multisite}
\end{align}
%\end{widetext}
where the first, second, third, and fourth terms comprise the effective bilinear two-site exchange interactions, %already introduced in Eq.~\eqref{eq:isoH}. 
%the only one present in the two-spin model.
the two-site four-spin (biquadratic) interactions, the three-site four-spin and the four-site four-spin interactions, respectively. All the summations in Eq.~\eqref{eq:4spin_multisite} are taken over mutually distinct indices, i.e., none of the different indices coincide. The multisite interaction parameters can be expressed in terms of the two-spin and four-spin integrals as,
\begin{equation}
    J_{ij}=-I_{ij}-I_{iiij}-I_{ijjj}+\sum_{k\notin\{i,j\}}\left(I_{ikjk}-2I_{ijkk} \right)
    \label{eq:J_multisite}
\end{equation}
\begin{align}
    B_{ij}=2I_{ijij}\;, \quad T_{ijk}=4I_{ikjk} \, ,
    \label{eq:B-T_multisite}
\end{align}
and
\begin{align}
K_{ijkl} = I_{ijkl}+I_{iklj}-I_{ikjl}\;.
\end{align}
Note that the overall minus sign
in the first term of Eq. (10) implies that positive/negative bilinear two-site interactions $J_{ij}$  refer to
ferromagnetic/antiferromagnetic pairwise couplings.

\section{Ab initio theory of local exchange interactions}  \label{sec:locint}
The method of infinitesimal rotations \cite{Liechtenstein1987local,ANTROPOV_1997} has been extended to non-collinear reference states by Szilva {\em et al.} \cite{Szilva2013} and later on used to explore the local DMI by Cardias \textit{et al.} \cite{cardias2020dzyaloshinskii,cardias2020first}. Here we give a short recap of the method within the multiple scattering formalism. There are two ways to implement infinitesimal rotations: either in terms of the single-site scattering matrices and the structural resolvents or in terms of the inverse of the $t$-matrices and the SPO matrix as done by Szilva {\em et al.} \cite{Szilva2013}.  As what follows, the formalism based on the single-site scattering matrices and the structural resolvents will be presented and in Appendix~\ref{appendix:pTau} it will be shown how the results can be transferred to the formalism of Szilva {\em et al.} \cite{Szilva2013}. %The details of the derivation can also be found in the Bachelor's thesis of Miklós Salánki \cite{salanki_bsc}. 

The interacting part of the grand potential for two-spin rotations is defined as,
\begin{equation}
    \delta \Omega_{ij} = \delta_{ij} \Omega -\delta_i \Omega -\delta_j \Omega \, ,
    \label{eq:Eint}
\end{equation}
where $\delta_i$ and $\delta_j$ denote variations due to the rotation of the spin at sites $i$ and $j$, respectively, while $\delta_{ij}$ stands for the variation by simultaneous rotations of the two spins. Based on Eq.~\eqref{eq:lloyd}, the single-spin variation of the grand potential can be calculated as
\begin{equation}
    \delta_i\Omega = \dfrac{1}{\pi}\Im \int_{-\infty}^{E_F} \dd \varepsilon \Tr \ln (\mathbf{I}-\delta\mathbf{t}^i\mathbf{G}) \, ,
    \label{eq:dOmegai}
\end{equation}
where the single-spin variation of the single-site scattering matrix is defined as 
\begin{equation}
\delta \mathbf{t}^i =\{ \underline{t}^i_s \, \delta \vec{e}_i \, \vec{\sigma} \; \delta_{in} \delta_{im}\} \, ,
\label{eq:dt-de}
\end{equation}
giving nonzero contribution only at site $i$. For infinitesimal rotation of the spin vector we have $\vec{e}_i\delta\vec{e}_i=0$, termed to as transverse fluctuation. In Eq.~\eqref{eq:dOmegai},
\begin{equation}
\mathbf{G} = \mathbf{G}_0 \left( \mathbf{I} - \mathbf{t} \mathbf{G}_0\right)^{-1} \, ,
\end{equation}
is the structural Green's function matrix of the system.

For the two-spin variation one can derive a similar expression,
\begin{equation}
     \delta_{ij}\Omega = \dfrac{1}{\pi}\Im \int_{-\infty}^{E_F} \dd \varepsilon \Tr \ln (\mathbf{I}-\delta\mathbf{t}^i\mathbf{G}-\delta\mathbf{t}^j\mathbf{G}) \, .
     \label{eq:dOmegaij1}
\end{equation}
Inserting Eqs.~\eqref{eq:dOmegai} and \eqref{eq:dOmegaij1} into Eq.~\eqref{eq:Eint} and expanding the logarithm as in \eqref{eq:ln1mx} to first order in the variation of the $t$-matrix we obtain,
\begin{align}
    \delta\Omega_{ij}&\simeq-\dfrac{1}{\pi} \Im \int_{-\infty}^{E_F} \dd \varepsilon \Tr (\delta \mathbf{t}^i \mathbf{G} \delta\mathbf{t}^j \mathbf{G} ) \\
    &= -\dfrac{1}{\pi} \Im \int_{-\infty}^{E_F} \dd \varepsilon \Tr ( \delta \matr{t}^i \matr{G}^{ij} \delta \matr{t}^j \matr{G}^{ji}) \, .
    \label{eq:dOmegaij2}
\end{align}
%where in Eq.~\eqref{eq:dOmegaij2} double underlines denote matrices with spin and angular momentum indices. 
We note that we arrived at an expression similar to \eqref{eq:Iij}, the Green's function matrix of the non-magnetic reference system being replaced by that of the magnetic system in the chosen spin-configuration. Since the Green's function is now spin dependent it can be decomposed in spin space using the Pauli matrices,
\begin{equation}
    \underline{\underline{G}}^{ij} = \underline{G}^{ij}_0 I_2 + \overrightarrow{\underline{G}}^{ij} \vec{\sigma} \, .
\end{equation}
Substituting this expression into \eqref{eq:dOmegaij2} and evaluating the trace in spin space we arrive at expressions, where quantities similar to the tensorial exchange interactions \cite{udvardi2003first} can be recognized. The two-spin variation of the grand potential separates into three contributions 
%according to the products of Pauli matrices the trace in spin trace had to be performed,
based on the number of Pauli matrices appearing in the spin trace,
\begin{equation}
    \delta \Omega_{ij} = \delta \Omega_{ij}^{(1)} +  \delta \Omega_{ij}^{(2)} + \delta \Omega_{ij}^{(3)} \, ,
    \label{eq:E123}
\end{equation}
where
\begin{align}
\delta \Omega_{ij}^{(1)} = & \: J_{ij} \delta\vec{e}_i \delta\vec{e}_j \, , \label{eq:loc_J}\\
\delta \Omega_{ij}^{(2)} = & \: \delta\vec{e}_i  \mathcal{S}_{ij} \delta\vec{e}_j \, ,
\label{eq:loc_anis}\\
\delta \Omega_{ij}^{(3)} = & \: \vec{D}_{ij} (\delta\vec{e}_i \times \delta\vec{e}_j)\;.
\label{eq:loc_dm}
\end{align}
The interaction parameters are defined as
\begin{equation}
    J_{ij}=A^{00}_{ij} - \sum_{\alpha\in\{x,y,z\}} A_{ij}^{\alpha\alpha} \, ,
    \label{eq:loc_J_tG}
\end{equation}
\begin{align}
    S_{ij}^{\alpha\beta}=A_{ij}^{\alpha\beta}+A_{ij}^{\beta\alpha} \, , %\quad  (\alpha,\beta\in\{x,y,z\}) \, ,
    \label{eq:loc_S_tG} 
\end{align}
where
\begin{equation}
    A_{ij}^{\mu\nu} = -\dfrac{2}{\pi}\Im \int_{-\infty}^{E_F} \dd \varepsilon \Tr_L \left( \underline{t}_s^i \underline{G}^{ij}_\mu  \underline{t}_s^j \underline{G}^{ji}_\nu \right)   \, ,
\end{equation}
$\mu,\nu \in \{0,x,y,z\}$ and
\begin{equation}
     D_{ij}^{\alpha}=-\dfrac{2}{\pi}\Re \int_{-\infty}^{E_F} \dd \varepsilon \Tr_L \left( \underline{t}_s^i \underline{G}^{ij}_0  \underline{t}_s^j \underline{G}^{ji}_\alpha-\underline{t}_s^i \underline{G}^{ij}_\alpha  \underline{t}_s^j \underline{G}^{ji}_0 \right) \;.
     \label{eq:loc_D_tG}
\end{equation}
%where $\alpha,\beta \in \{x,y,z\}$. 
The above expressions will be referenced as the $tG$ formalism of the local magnetic interactions.

Similar equations can be obtained for the local exchange interactions if we switch to the formalism used by Antropov {\em et al.} \cite{ANTROPOV_1997}, Szilva {\em et al.} \cite{Szilva2013} and Cardias {\em et al.} \cite{cardias2020dzyaloshinskii,cardias2020first}, with  the $t$  and $G$ matrices being replaced by the $p=t^{-1}$ and $\tau$ matrices, respectively. Hence, we will reference it as the $p\tau$ formalism of local magnetic interactions. The details can be found in Appendix \ref{appendix:pTau}. From the derivation presented there, it appears that the $tG$ formula \eqref{eq:dOmegaij2} and the $p\tau$ formula \eqref{eq:dOmegaij-pt} are identical for infinitesimal rotations of the spin vectors. %This is clearly confirmed by our numerical results, see Fig.~\ref{fig:dEij_tG_pT}.
However, there is no guarantee that the three contributions to $\delta \Omega_{ij}$ are identical in the two local formalisms. This will be illustrated by numerical calculations in the next section.
%since longitudinal contributions, although vanishing when summed, can give rise to uncontrolled differences between them. 

%One has to be very careful when interpreting the results of this section. 
It should also be emphasized that relativistic effects like spin-orbit coupling (SOC) were excluded from the theory, but we obtained interactions, like the Dzyaloshinskii--Moriya interaction or exchange anisotropy, known to be related to the SOC. %This result is very confusing and can lead to the misinterpretation of the local interactions as discussed in Ref.~\cite{dos2021proper}. 
The origin of the DMI in the absence of SOC was explained in Ref.~\cite{dos2021proper} 
%and \cite{Laszloffy_dissertation} 
demonstrating that the two-spin variation of the isotropic four-spin multisite Hamiltonian in \eqref{eq:4spin_multisite} leads to local interactions which appear in the same form as \eqref{eq:loc_anis} and \eqref{eq:loc_dm}. %In the next section we present calculations for the local interactions, together with a direct comparison of the interactions entering the local and global spin models. 

\section{Local interactions for a Cr trimer on Au(111)}
For our numerical considerations we choose an equilateral Cr trimer on the (111) surface of Au, used in many previous studies \cite{antal2008first,cardias2020first,dos2021proper,martinez2023relativistic} and we compare the numerical results obtained for the $tG$ and $p\tau$ types of local interactions implemented within the embedded cluster multiple scattering Green's-function framework \cite{lazarovits2002fully}.

\begin{figure}[t!]
    \centering
    \includegraphics[width=.6\linewidth]{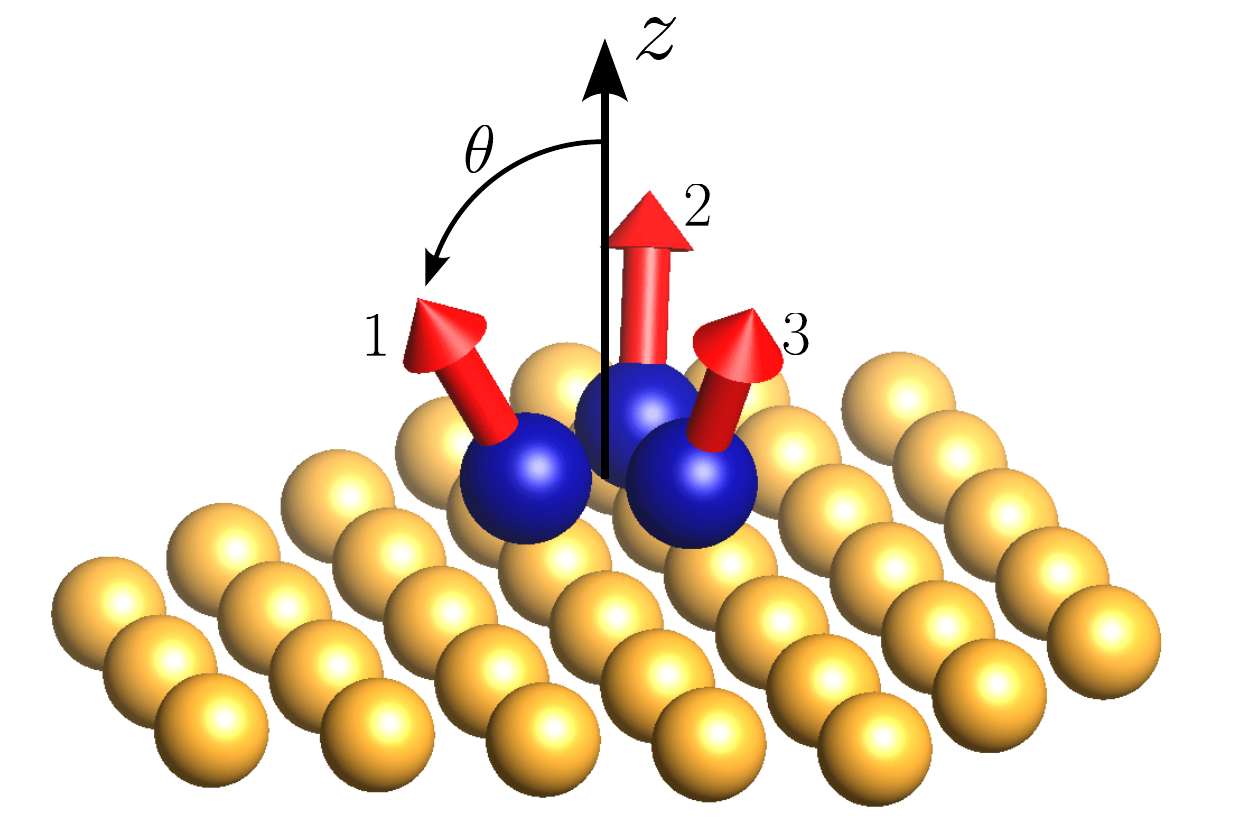}
    \caption{The spin configuration of the Cr trimer on the surface of Au(111) used for the calculation of the local interactions.}
    \label{fig:cr3au111}
\end{figure}

The Cr trimer on the surface of Au(111), in short Cr$_3$/Au(111), is considered in a non-relaxed geometry, where the Au and the Cr atoms occupy the positions in an ideal face-centered cubic lattice, with lattice constant $287.4$~pm. The Cr atoms are sitting above the hollow positions of the top Au layer forming a right equilateral triangle. The host is calculated with 16 energy points on a semicircular path on the upper complex half plane with 2160 $k$ points in the irreducible wedge of the 2D Brillouin zone at the Fermi level and an angular momentum cut-off $\ell_{\max}=2$. The local interactions are calculated along a path in the configuration space illustrated in Fig.~\ref{fig:cr3au111}, where $\vartheta$ was changed in $10^\circ$ steps using the self-consistent potential obtained for each value of $\vartheta$.
We have verified that the magnitudes of the magnetic moments remain constant within a relative accuracy of 1~\% along the path considered, providing strong support for the rigid-moment approximation employed throughout our work.

\begin{figure}[htb]
    \centering
    \includegraphics[width=.98\linewidth]{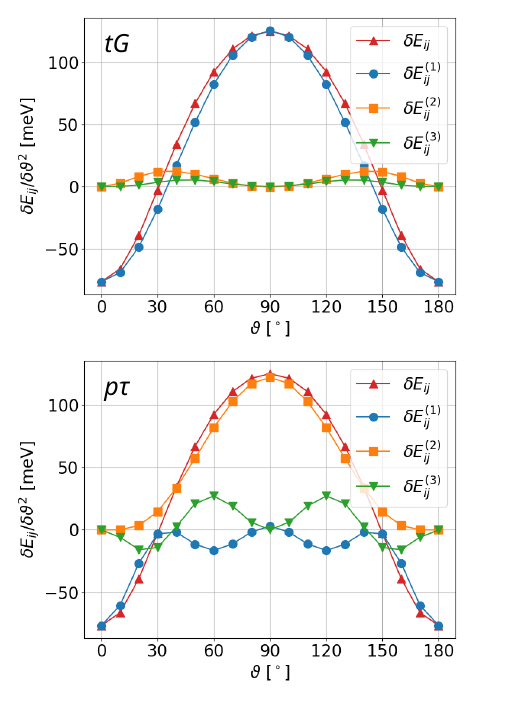}
    \caption{The two-spin variation of the energy for Cr$_3$/Au(111) along the path shown in Fig.~\ref{fig:cr3au111}, by varying the $\vartheta$ angle of the spin vectors: \textbf{(a)} $tG$ formalism and  \textbf{(b)}  $p\tau$ formalism.}
    \label{fig:dEij_tG_pT}
\end{figure}

The two-spin variation of the grand potential \eqref{eq:E123}, henceforth referred to as the energy $E$ of the spin system, and its three contributions are shown in Fig.~\ref{fig:dEij_tG_pT} along the chosen path as calculated from both the $tG$ and $p\tau$ formalisms. Note that $\delta E_{ij}$ is normalized to unit variation of the angle $\vartheta$, i.e., $\delta E_{ij}/\delta\vartheta^2$ is presented in the figure. As predicted in the previous section, the total variations (red curves) agree perfectly for the two formalisms. The decompositions of the two-spin variation of the energy into the three contributions are, however, significantly different in the two types of calculations. For example, around $\vartheta = 90^\circ$, $\delta E_{ij}^{(1)}$ provides the largest contribution to $\delta E_{ij}$ in the $tG$ formalism, while $\delta E_{ij}^{(2)}$ and $\delta E_{ij}^{(3)}$ approach zero. In contrast, in the $p\tau$ formalism, $\delta E_{ij}^{(2)}$ dominates $\delta E_{ij}$, whereas $\delta E_{ij}^{(1)}$ and $\delta E_{ij}^{(3)}$ are suppressed.

The different contributions to the energy variation can further be analyzed in terms of the interaction parameters. The interactions in \eqref{eq:loc_J_tG}, \eqref{eq:loc_S_tG} and \eqref{eq:loc_D_tG}, or their analogues in the $p\tau$ formalism can be organized into a $3\times 3$ matrix, similarly as %in \eqref{eq:J_decomp}
one introduces the exchange tensor for the relativistic two-spin interactions \cite{udvardi2003first}. Then the number of independent interaction parameters can be reduced based on symmetry arguments in the same way as in Ref.~\cite{nyari2019weak}. The simplest matrix structure is assigned with the interaction between the two sites at the lower edge of the triangle labeled by 13 (the bond along the $x$ axis),
%{\footnotesize 
\begin{equation}
\mathcal{J}_{13}=
\left(
\begin{matrix}
{J}+S_{x} & D_\perp & - D_\parallel \\
-D_\perp & {J}+S_y & S_{yz} \\
D_\parallel & S_{yz} & {J}-S_{x}-S_{y}
\end{matrix}
\right) \, .
\label{eq:j13_general}
\end{equation}
%}

\begin{figure*}[t!]
    \centering
    \includegraphics[width=.9\linewidth]{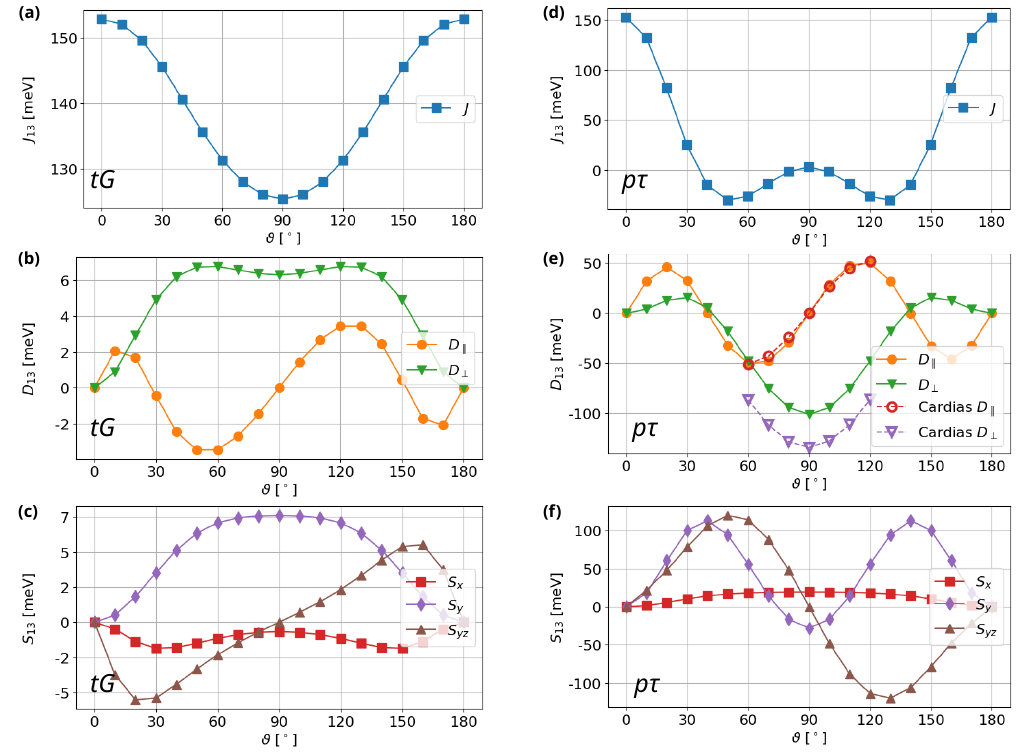}
    \caption{The local interaction parameters for the Cr$_3$/Au(111) as specified in the matrix \eqref{eq:j13_general} for the $tG$ formalism (left panels) and for the $p\tau$ formalism (right panels). The data labeled as `Cardias' in panel \textbf{(e)} is copied from Ref.~\cite{cardias2020dzyaloshinskii}.}
    \label{fig:loc-int}
\end{figure*}

The interaction matrices related to the other two pairs can be obtained by rotation %and transposition 
of this matrix, which means that the local interactions are described by the six parameters in Eq.~\eqref{eq:j13_general}. We should note that there exists a distinct set of these parameters for each spin configuration considered. The interaction parameters for Cr$_3$/Au(111) are plotted in Fig.~\ref{fig:loc-int}. It is evident that both the energy contributions and the interaction parameters differ between the two formalisms. Our calculations then suggest that neither the decomposition of the energy variations nor the corresponding interactions like local DMI can uniquely be defined. For the collinear case ($\vartheta=0$) the local DMI and the matrix elements of $\mathcal{S}_{ij}$ vanish, and only the isotropic interaction differs from zero. Interestingly, in this limit the value of $J_{ij}$ is the same for the formalism $tG$ and $p\tau$, representing the well-known Heisenberg exchange coupling invented by Liechtenstein {\em et al.} \cite{Liechtenstein1987local}. 

Despite the ambiguity of the local interaction parameters, a comparison can be made between parameters that are calculated within the same formalism. Fig.~\ref{fig:loc-int}(e) compares the components of the local DM vector we calculated in the $p\tau$ formalism and those by Cardias {\em et al.} in Ref.~\cite{cardias2020dzyaloshinskii}. Apparently, there is an almost perfect agreement between the parallel components, and the perpendicular components are also in the same range of magnitude, contrary to the results shown in Fig.~2 of Ref.~\cite{dos2021proper}. 

\begin{comment}
This surprising result becomes clear when we analyze the transition between the two formalisms and identify terms that vanish in the energy variation, yet yield finite interaction contributions. The details of this analysis is presented in Appendix \ref{appendix:diff_tG_pTau}. The energy variation proved to be identical when calculated from the two formalisms, but the local interaction matrices are different. The term, which gives rise to the difference in the local interaction matrices, is proportional to the spin vectors, related to longitudinal energy variations. Since the method of infinitesimal rotations relies on transverse variations only, such terms should not occur in the interaction matrices. %Consequently, these longitudinal terms has to be projected out to restrict the interactions to the transverse subspace only.
\end{comment}

\section{Local interactions from the global spin model}

In the previous section we concluded that the method of infinitesimal rotations as described in Section~\ref{sec:locint} %implemented by Szilva {\em et al.} \cite{Szilva2013} and Cardias {\em et al.} \cite{cardias2020dzyaloshinskii,cardias2020first} 
doesn't provide unique local interactions. In order to get a deeper understanding why this ambiguity happens, we make use of the global isotropic four-spin model introduced in Sec.~\ref{sec:glob4spin} and analytically derive the two-spin variation of the energy \cite{LaszloffyPhd}. %following the calculations of András Lászlóffy\cite{Laszloffy_dissertation}. 

The multisite Hamiltonian defined in Eq.~\eqref{eq:4spin_multisite} applied to an equilateral trimer reads as
\begin{align}
    H=-&J(\vec{e}_1\vec{e}_2+\vec{e}_1\vec{e}_3+\vec{e}_2\vec{e}_3) \nonumber \\+&\dfrac{1}{2} B\left( (\vec{e}_1\vec{e}_2)^2 + (\vec{e}_1\vec{e}_3)^2 + (\vec{e}_2\vec{e}_3)^2 \right) \nonumber \\
    +& \dfrac{1}{2}T \left(  (\vec{e}_1\vec{e}_2)(\vec{e}_1\vec{e}_3) + (\vec{e}_1\vec{e}_2)(\vec{e}_3\vec{e}_2) + (\vec{e}_1\vec{e}_3)(\vec{e}_2\vec{e}_3) \right) \, ,
    \label{eq:fourspinham-trimer}
\end{align}
that means we have only three independent parameters, $J$, $B$, and $T$.
The two-spin variation of this Hamiltonian with respect to spins $i$ and $j$ can be expressed as

\begin{small}
\begin{align}
\delta E_{ij}&=\left[ -J+B(\vec{e}_i\vec{e}_j)\right] (\delta\vec{e}_i\delta\vec{e}_j)\nonumber \\
&+\dfrac{1}{2} \left[ T\left( (\vec{e}_i\vec{e}_k)+ (\vec{e}_j\vec{e}_k)\right)\right]\delta\vec{e}_i \delta\vec{e}_j \nonumber \\
&+ B(\vec{e}_j\delta\vec{e}_i)(\vec{e}_i\delta\vec{e}_j) \nonumber \\
&+\dfrac{1}{2} T \left(  (\vec{e}_k\delta\vec{e}_i)(\vec{e}_i\delta\vec{e}_j)+(\vec{e}_j\delta\vec{e}_i)(\vec{e}_k\delta\vec{e}_j)+(\vec{e}_k\delta\vec{e}_i)(\vec{e}_k\delta\vec{e}_j) \right) \, ,
\label{eq:local-deltaE_ij}
\end{align}
\end{small}

\noindent
where $i,j,k\in\{1,2,3 \}$ and $i\neq j \neq k$. This can be written as

\begin{equation}
\delta E_{ij}=\delta \vec{e}_i \mathcal{J}_{ij} \delta \vec{e}_j
\label{eq:dEij}
\end{equation}

\noindent
with the local interaction matrix,

\begin{align}
\mathcal{J}_{ij}=&\left[ -J+B(\vec{e}_i\vec{e}_j)+\dfrac{1}{2}T\left( (\vec{e}_i\vec{e}_k)+ (\vec{e}_j\vec{e}_k)\right)\right]\mathcal{I}_{3} \nonumber \\ &+ B(\vec{e}_j\circ\vec{e}_i) 
+\dfrac{1}{2} T \left(  (\vec{e}_k\circ\vec{e}_i)+(\vec{e}_j\circ\vec{e}_k)+(\vec{e}_k\circ\vec{e}_k) \right)\;.
\label{eq:4spin-local}
\end{align}

Substituting the spin configuration of the Cr trimer indicated in Fig.~\ref{fig:cr3au111}, 

\begin{footnotesize}
\begin{align}
\vec{e}_1=\left( \begin{matrix} -\frac{\sqrt3}{2}\sin\vartheta \\ -\frac{1}{2} \sin\vartheta \\ \cos\vartheta \end{matrix} \right)  
\quad 
\vec{e}_2=\left( \begin{matrix} 0 \\  \sin\vartheta \\ \cos\vartheta \end{matrix} \right)
\quad
\vec{e}_3=\left( \begin{matrix} \frac{\sqrt3}{2}\sin\vartheta \\ -\frac{1}{2} \sin\vartheta \\ \cos\vartheta \end{matrix} \right)
\, ,
\label{eq:local-spin-vectors}
\end{align}
\end{footnotesize}

\noindent
into expression \eqref{eq:4spin-local}, the explicit form of the interaction can be given as a function of $\vartheta$. The local interaction matrix between spins 1 and 3 then reads as %which has the simplest form,
\begin{widetext}
\begin{footnotesize}
\begin{equation}
\mathcal{J}_{13}= 
\left(\begin{matrix}
 - J + B + T -\left( \dfrac{9}{4} B + \dfrac{3}{2} T \right)\sin^{2}\vartheta  & 
-\dfrac{\sqrt{3}}{4} (B - T) \sin^{2}\vartheta & 
\dfrac{\sqrt{3}}{4} \left(B + \dfrac{T}{2}\right) \sin2\vartheta\\
\dfrac{\sqrt{3}}{4} (B - T) \sin^{2}\vartheta &
 - J + B + T - \left(\dfrac{5}{4} B + \dfrac{3}{2} T \right)\sin^{2}\vartheta &
-\dfrac{1}{4}\left( B - \dfrac{3}{2} T\right) \sin2\vartheta\\
-\dfrac{\sqrt{3}}{4} \left(B + \dfrac{T}{2}\right) \sin2\vartheta &
-\dfrac{1}{4}\left( B - \dfrac{3}{2} T\right) \sin2\vartheta &
 - J + 2 B + \dfrac{5}{2} T - \left(\dfrac{5}{2} B + 3 T\right) \sin^{2}\vartheta 
\end{matrix}
\right) \, .
\label{eq:jij_orig}
\end{equation}
\end{footnotesize}
\end{widetext}

\begin{figure}
    \centering
    \includegraphics[width=0.85\linewidth]{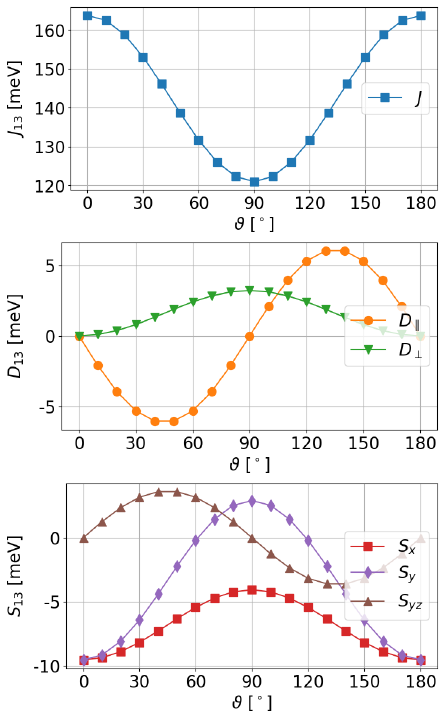}
    \caption{The local interaction parameters for the Cr$_3$/Au(111) from the global four-spin model as defined in Eqs.~\eqref{eq:jpar-glob}--\eqref{eq:syz-glob}. The parameters entering these relations were calculated based on expressions \eqref{eq:J_multisite} and \eqref{eq:B-T_multisite}: $J=132.86$~meV, $B=6.97$~meV and $T=14.41$~meV.}
    \label{fig:loc-int-from-global}
\end{figure}

\noindent
%where the antisymmetric part of the matrix is obviously related to the DM vector. % derived in Ref.~\cite{Laszloffy_dissertation}. At this point we diverge from the derivation of Lászlóffy\cite{Laszloffy_dissertation} 
Note that the above matrix matches the form of $\mathcal{J}_{13}$ in Eq.~\eqref{eq:j13_general} dictated by symmetry with the following parameters: 
\begin{equation}
J(\vartheta) = - J + \frac{4}{3} B + \frac{3}{2} T -  2 \left( B + T \right) \sin^2 \vartheta \, , 
\label{eq:jpar-glob} 
\end{equation}
\begin{equation}
D_\perp(\vartheta) =  -\dfrac{\sqrt{3}}{4} (B - T) \sin^{2}\vartheta \, , 
\label{eq:dperp-glob} 
\end{equation}
\begin{equation}
D_\parallel(\vartheta) =  - \dfrac{\sqrt{3}}{4} \left(B + \dfrac{T}{2}\right) \sin2\vartheta \, , 
\label{eq:dpar-glob} 
\end{equation}
\begin{equation}
S_x (\vartheta) =  - \dfrac{1}{3} B -\frac{1}{2} T -\left( \dfrac{1}{4} B - \dfrac{1}{2} T \right)\sin^{2}\vartheta \, , 
\label{eq:sx-glob} 
\end{equation}
\begin{equation}
S_y (\vartheta) = - \dfrac{1}{3} B -\frac{1}{2} T + \left( \dfrac{3}{4} B + \dfrac{1}{2} T \right)\sin^{2}\vartheta \, , 
\label{eq:sy-glob} 
\end{equation}
\begin{equation}
S_{yz} (\vartheta) = -\dfrac{1}{4}\left( B - \dfrac{3}{2} T\right) \sin2\vartheta \, .
\label{eq:syz-glob}
\end{equation}
The local interactions obtained from the expressions \eqref{eq:jpar-glob}--\eqref{eq:syz-glob} are presented in Fig.~\ref{fig:loc-int-from-global}. Apparently, the energy range of these parameters compares well to that of the parameters calculated from the $tG$ local model, see left panels of Fig.~\ref{fig:loc-int}. This can be readily understood, as the two calculations rely on closely related formalisms. Nevertheless, apart from the isotropic parameter $J_{13}$, the interactions obtained from the $tG$ local model show a more complicated $\vartheta$--dependence than those from the global model. %Beyond the fact that the two local formalisms yield remarkably different local interaction parameters, this disagreement poses another puzzle to solve. 
An even more serious problem arises from the fact that adding longitudinal terms, proportional to $\vec{e}_i\delta\vec{e}_i$ or $\vec{e}_j\delta\vec{e}_j$, that leave $\delta E_{ij}$ invariant, can arbitrarily modify the local interaction parameters. An example is presented in Appendix \ref{appendix:locparamb}.  
%Although this difference could, in principle, be attributed to the absence of higher-order (beyond fourth-order) interactions in the global model \eqref{eq:fourspinham-trimer}, it is almost fully resolved by a consistent treatment of classical spin rotations.

\section{Projection to the transverse subspace}

In this section, we demonstrate that all issues raised above are resolved by systematically removing the longitudinal contributions from the two-spin energy variation \eqref{eq:dEij}. Let's define the basis vectors of the local frame of reference at site $i$ as

%\subsection{Analytic expressions based on the global model}
%In a classical spin model, the spins are considered to be unit vectors, and the energy depends only on the orientation of the spins. 
%Given the spin vector at site $i$ in spherical coordinates,

\begin{equation}
\vec{e}^{\: r}_i \equiv \vec{e}_i=
\left(
\begin{matrix}
\sin\vartheta_i \cos\phi_i \\
\sin\vartheta_i \sin\phi_i \\
\cos\vartheta_i 
\end{matrix}
\right) \, ,
\end{equation}

\begin{equation}
\vec{e}_i^{\;\vartheta} \equiv \frac{d \vec{e}_i}{d \vartheta_i} =
\left(
\begin{matrix}
\cos\vartheta_i \cos\phi_i \\
\cos\vartheta_i \sin\phi_i \\
-\sin\vartheta_i 
\end{matrix}
\right) \, ,
\end{equation}

\begin{equation}
\vec{e}_i^{\;\phi} \equiv \frac{d \vec{e}_i}{d \phi_i} =
\left(
\begin{matrix}
-\sin\phi_i \\
\cos\phi_i \\
0
\end{matrix}
\right)\;,
\end{equation}

\noindent
and the matrix elements of $\mathcal{J}_{ij}$  with respect to these basis vectors,
\begin{equation}
    J_{ij}^{\mu\nu}=\vec{e}^{\;\mu}_i \mathcal{J}_{ij} \vec{e}^{\;\nu}_j\;,
\end{equation}

\noindent
with $\mu,\nu \in \{r, \vartheta, \phi \}$. The interaction matrix can then be expressed as 

\begin{equation}
\mathcal{J}_{ij}=\sum_{\mu,\nu\in\{r, \vartheta,\phi\}} J_{ij}^{\mu\nu} \: \left( \vec{e}^{\;\mu}_i \circ  \vec{e}^{\;\nu}_j \right) \; ,
\label{eq:jij-localframe}
\end{equation}

\noindent
where $\vec{a} \circ \vec{b}$ denotes the dyadic product of the vectors $\vec{a}$ and $\vec{b}$.

The variation of the spin vector by infinitesimal rotation can be written as

\begin{equation}
    \delta\vec{e}_i=\delta\vartheta_i \vec{e}_i^{\;\vartheta}+\delta\phi_i\sin\vartheta_i \vec{e}_i^{\;\phi}\;.
    \label{eq:evec-rot}
\end{equation}

\noindent
The two-spin transverse variation of the energy is then given by

\begin{align}
\delta E_{ij}&=
\delta\vartheta_i\delta\vartheta_j \, J_{ij}^{\vartheta\vartheta} +
\delta\phi_i\delta\phi_j \, \sin\vartheta_i \sin\vartheta_j \, J_{ij}^{\phi\phi} \nonumber \\ &+
\delta\vartheta_i\delta\phi_j \, \sin\vartheta_j \, J_{ij}^{\vartheta\phi} +
\delta\phi_i\delta\vartheta_j \, \sin\vartheta_i  \, J_{ij}^{\phi\vartheta}\;,
\label{eq:dEij-loc}
\end{align}

\noindent
from which the transverse second derivatives of the energy can be inferred,
\begin{align}
    \dfrac{\delta^2E_{ij}}{\delta\vartheta_i\delta\vartheta_j}=J_{ij}^{\vartheta\vartheta} \;, \quad 
    \dfrac{\delta^2E_{ij}}{\delta\phi_i\delta\phi_j}=J_{ij}^{\phi\phi} \sin\vartheta_i \sin\vartheta_j \; , 
    \label{eq:loc-der-1} \\ 
    \dfrac{\delta^2E_{ij}}{\delta\vartheta_i\delta\phi_j}=J_{ij}^{\vartheta\phi} \sin\vartheta_j \;, \quad 
    \dfrac{\delta^2E_{ij}}{\delta\phi_i\delta\vartheta_j}=J_{ij}^{\phi\vartheta} \sin\vartheta_i \;. \label{eq:loc-der-2}
\end{align}
%Note that, when one would allow the change of the spin vectors' magnitude $r$, the longitudinal derivatives can also be calculated by %defining $\vec{e}_i^{\: r} \equiv \vec{e}_i$,
% \begin{align}
%     \dfrac{\delta^2E_{ij}}{\delta r_i\delta\phi_j}=J_{ij}^{r \phi} \sin\vartheta_j \; , \quad
%     \dfrac{\delta^2E_{ij}}{\delta\phi_i \delta r_j}=J_{ij}^{\phi r} \sin\vartheta_i \; ,
%     \label{eq:loc-der-r1} \\ 
%     \dfrac{\delta^2E_{ij}}{\delta r_i \delta\vartheta_j}=J_{ij}^{r \vartheta} \;, \quad 
%     \dfrac{\delta^2E_{ij}}{\delta\vartheta_i \delta r_j}=J_{ij}^{\vartheta r} \;, \quad 
%     \dfrac{\delta^2E_{ij}}{\delta r_i\delta r_j}=J_{ij}^{rr} \;.  
%     \label{eq:loc-der-r2}
% \end{align}

Since in Eq.~\eqref{eq:dEij-loc} only transverse matrix elements of $\mathcal{J}_{ij}$ are present, a unique definition of the local interaction matrix can be achieved by projecting it to the transverse subspace,

\begin{equation}
\widehat{\mathcal{J}}_{ij}=\sum_{\mu,\nu\in\{\vartheta,\phi\}} J_{ij}^{\mu\nu} \; \left( \vec{e}^{\;\mu}_i \circ  \vec{e}^{\;\nu}_j \right) \; .
\label{eq:proj}
\end{equation}
%where the notation %\ $\; \widehat{\ } \;$ \ refers to the projected quantities. 
This matrix can be used in the two-spin variation of the energy \eqref{eq:dEij} when only transverse variations (infinitesimal rotations) of the spin vectors \eqref{eq:evec-rot} are considered, while the longitudinal components of the matrix \eqref{eq:jij-localframe} are irrelevant.

For the pair 1--3, the local projected interaction matrix can be written in the form defined in Eq.~\eqref{eq:j13_general} with the `isotropic' parameter,
\begin{widetext}
\begin{equation}
\widehat{J}(\vartheta) = -\dfrac{2}{3}\left(J-B-T \right)+\left(J-3B-\dfrac{9}{4}T\right) \sin^2\vartheta - \left(\dfrac{3}{4}J-\dfrac{9}{2}B-\dfrac{33}{8}T \right)\sin^4\vartheta - \left(\dfrac{9}{4}\ B+\dfrac{45}{16}T\right)\sin^6\vartheta \, ,
\label{eq:jhat}
\end{equation}
the `DM-like' parameters,

\begin{align}
\widehat{D}_\perp(\vartheta)&=\dfrac{\sqrt{3}}{32}\left[ -8(J-B)+44T+(12J-48B-84T)\sin^2\vartheta+(36B+45T)\sin^4\vartheta\right]\sin^2\vartheta \, , \label{eq:dperphat}  \\
\widehat{D}_\parallel(\vartheta)&=\dfrac{\sqrt{3}}{32}\left[ -8(J-B-T)+(12J-48B-48T)\sin^2\vartheta+(36B+45T)\sin^4\vartheta\right]\sin2\vartheta \, ,\label{eq:dparhat} 
\end{align}
and the `anisotropy' parameters,
\begin{align}
&\widehat{S}_{x}(\vartheta)=-\dfrac{1}{3}(J-B-T)+\dfrac{1}{8}(10J-30B-27T)\sin^2\vartheta-\dfrac{1}{8}(3J-36B-39T)\sin^4\vartheta-\frac{1}{32}(36B+45T)\sin^6\vartheta \, , \label{eq:sxhat} \\
&\widehat{S}_{y}(\vartheta)=-\dfrac{1}{3}(J-B-T)-\dfrac{1}{8}(6J-10B-13T)\sin^2\vartheta+\dfrac{1}{8}(9J-36B-51T)\sin^4\vartheta+\frac{1}{32}(108B+135T)\sin^6\vartheta \, , \label{eq:syhat} \\
&\widehat{S}_{yz}(\vartheta)=\dfrac{1}{8}\left[ -8(J-B-T)-(12J-36T)\sin^2\vartheta-(36B+45T)\sin^4\vartheta\right]\sin2\vartheta\, .
\label{eq:syzhat}
\end{align}
\end{widetext}
By comparing the above projected interactions with the original ones in Eq.~\eqref{eq:jpar-glob}--\eqref{eq:syz-glob}, one can see that the projection introduces higher-order harmonics in $\vartheta$. 

%\subsection{Numerical results}
%Previously we showed that during the transition from the $tG$ to the $p\tau$ formalism we neglected the second term in \eqref{eq:pdtp_full} which is proportional to the spin vectors and, therefore, gives zero contribution to the transverse derivatives. This fact can also be verified numerically by 

\begin{figure}[htb!]
    \centering
    \includegraphics[width=1.0\linewidth]{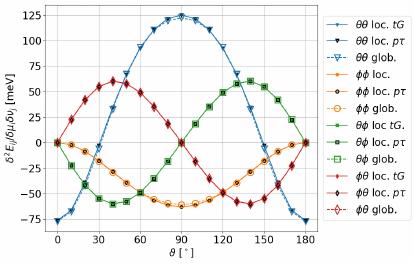}
    \caption{The transverse two-spin derivatives of the energy for the Cr pair 1--3 in Cr$_3$/Au(111) along the path sketched in Fig.~\ref{fig:cr3au111} as calculated using the two different local spin models and the global spin model. %: \textbf{(a)} transverse derivatives. \textbf{(b)} longitudinal derivatives. The derivatives are given in the corresponding local coordinate systems.
    }
    \label{fig:derivatives}
\end{figure}

The transverse two-spin derivatives of the energy, Eqs.~\eqref{eq:loc-der-1} and \eqref{eq:loc-der-2} can also be evaluated with the local magnetic interactions, see Eqs.~\eqref{eq:loc_J_tG}--\eqref{eq:loc_D_tG}. In Fig.~\ref{fig:derivatives} we show the calculated transverse two-spin derivatives for pair 1--3 along the chosen path in the configuration space using the $tG$ and the $p\tau$ formalisms, as well as the global four-spin model shown in \eqref{eq:4spin-local}. From Fig.~\ref{fig:derivatives} one can infer that the transverse second derivatives based on the two local models almost perfectly agree (in fact, the difference is below 1 $\mu$eV). They are also in very good agreement with those obtained from the global four-spin model, with a difference of about 1 meV which still refers to a relative accuracy of 1 \%.
This small difference likely arises due to higher-order multispin interactions incorporated in the local models, but not present in the global spin model \eqref{eq:4spin_iso_ham}.

% In Fig.~\ref{fig:derivatives}(b) the derivatives containing longitudinal components show a different picture: none of the models compare well. This verifies that all the differences appearing in Fig.~\ref{fig:loc-int} originate from the longitudinal components of the interaction matrix. %It has to be noted that comparing the $tG$ and the global model only the $rr$ derivative differs considerably. This is related to the fact that they rely on the same formalism using the $t$ and $G$ matrices. Laci: ide kellene a Dyson-egyenletes analízis. Anélkül nem igazán erőltetném ezt az összehasonlítást.

\begin{figure*}[htb!]
    \centering
    \includegraphics[width=0.9\linewidth,trim=0cm 0cm 0cm 0cm,clip]{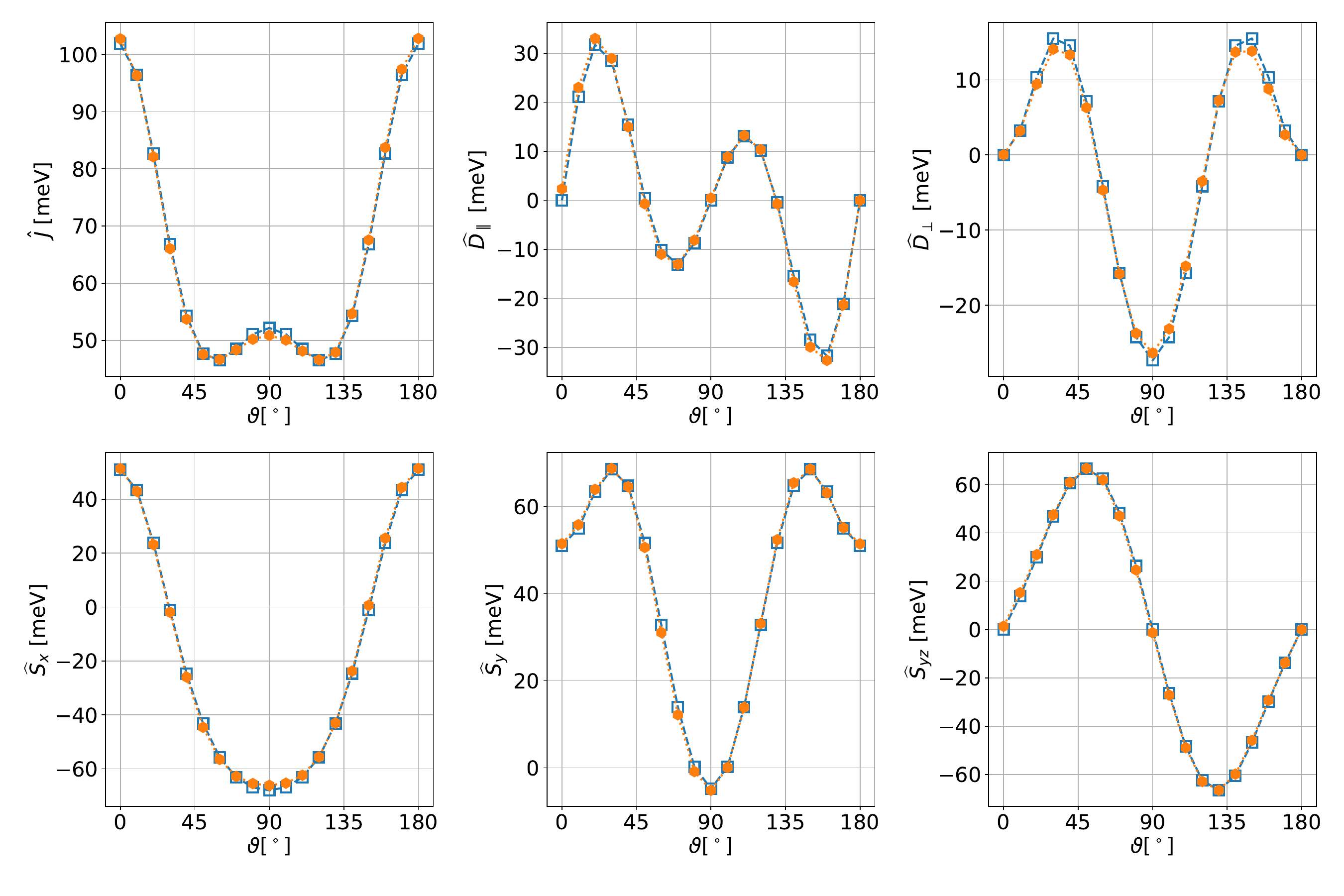}
    \caption{The local projected interaction parameters for Cr$_3$/Au(111) calculated from the local spin model in the $tG$ formalism (blue open squares) and from the global four-spin model by using the analytic expressions \eqref{eq:jhat}-\eqref{eq:syzhat} (orange solid hexagons). Dashed and dotted lines serve as guide for the eyes.}
    \label{fig:loc-vs-glob-int}
\end{figure*}

Finally, in this section, we calculate the local interaction matrices projected onto the transverse subspace [Eq.~\eqref{eq:proj}] for both the local and global spin models. As demonstrated in Fig.~\ref{fig:derivatives}, the transverse second derivatives do not depend on which local formalism is used, therefore, the matrix elements of $\widehat{\mathcal{J}}_{13}$ are also independent on the type of the formalism.  %by reading off the parameters in \eqref{eq:j13_general} from the matrix elements. 
 %In Fig.~\ref{fig:derivatives} we confirmed that the difference of the $tG$ and $p\tau$ formalism is coming from the different longitudinal components introduced by the transition between the two models. Since the longitudinal derivatives are only an artifact of the calculation -- and irrelevant for constructing a spin model -- the two methods obviously lead to the same model, so from now on we consider only the transverse parameters calculated in the $tG$ formalism.
In Fig.~\ref{fig:loc-vs-glob-int}, we compare the projected interaction parameters calculated from the local spin model with the analytic expressions obtained from the global isotropic four-spin model. Apparently, for each of the six parameters, the two curves agree with high accuracy, which is not surprising given the excellent agreement found for the corresponding transverse derivatives (see Fig.~\ref{fig:derivatives}). 
%The effect of the projection on the interaction parameters is best demonstrated when Fig.~\ref{fig:loc-vs-glob-int} is compared with Fig.~\ref{fig:loc-int}(a),(b) and (c), where the unprojected local interactions are plotted for the $tG$ formalism. %As clear from the analytic formulas \eqref{eq:jhat}--\eqref{eq:syzhat}, the projected interaction parameters show a complex $\vartheta$ dependence. 

\section{Fe and Cr dimers on Cu(001)}
\begin{figure}[htb!]
    \centering
    \includegraphics[width=1.0\linewidth,trim= 0cm 0cm 0cm 0cm]{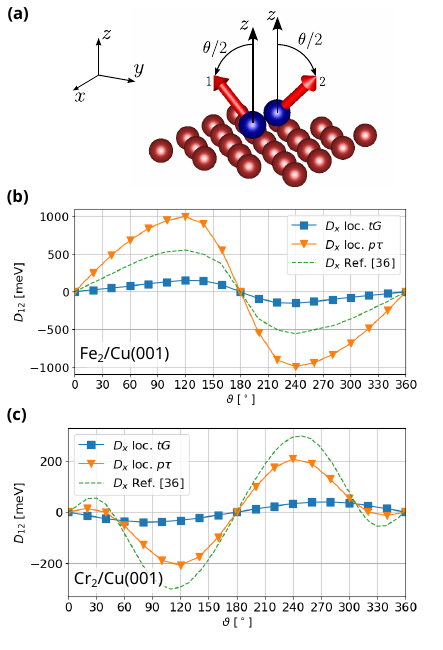}
    \caption{\textbf{(a)} Illustration of the spin-configuration used for the dimers on Cu(001). The spin vectors are confined to the $yz$ plane.  \textbf{(b)} and \textbf{(c)} Calculated local DMI parameters for Fe$_2$/Cu(001) and Cr$_2$/Cu(001), respectively. Note that, by symmetry, the local DM vector is parallel to the M–M bond (M = Fe or Cr).
    }
    \label{fig:cu}
\end{figure}

So far, using the textbook example of a Cr trimer on Au(111), we demonstrated in detail how the two formalisms for describing local interactions are related and how local, configuration-dependent interactions can be expressed in terms of global, configuration-independent parameters. To prove the robustness of the method, we also considered dimers of Fe and Cr on the (001) surface of Cu as studied by Cardias \textit{et al.} \cite{cardias2023unraveling}. The dimer is placed in nearest-neighbor positions along the (100) direction on the surface, with no structural relaxation performed. For the configuration path we choose a symmetric tilting of the spins in the plane defined by the bond of the dimer with a relative angle $\vartheta$ as illustrated in Fig.~\ref{fig:cu}(a). 

In Ref.~\cite{cardias2023unraveling}, the local DMI for dimers, including Fe and Cr on Cu(001), was reported both in the presence and absence of SOC using what we refer to as the $p\tau$ formalism. In Fig.~\ref{fig:cu}(b) and (c) we compare our results with the ones shown in Fig.~3 of Ref.~\cite{cardias2023unraveling} for the $x$ component of the DMI vector in the absence of SOC. It should be noted that the two atoms of the dimer are related by mirror symmetry, implying that the (non-relativistic) local DM vector is parallel to the line connecting them. The results obtained from the $p\tau$ formalism show good qualitative agreement with those of Ref.~\cite{cardias2023unraveling}: %with much larger values of the interaction parameters compared to the $tG$ formalism, 
for the Fe dimer our values are approximately twice as large, while for the Cr dimer we found a scaling factor of about $2/3$ compared to those in Ref. \cite{cardias2023unraveling}. However the results from the $tG$ formalism significantly differ both qualitatively and quantitatively. This again confirms the ambiguity in the calculation of the local DMI. %The quantities with physical meaning again the transverse second derivatives of the energy. Furthermore one should take into account all the parameters of the local interaction matrix next to the DMI term to obtain the relevant physical quantities. 

%To show that, how the local framework simplifies in the context of global interaction parameters, we derive the relation of the global multi-spin model and the local model. 
Next, we attempted to establish the relationship between the local model parameters and those of the global spin Hamiltonian. %For that reason we slightly modify the spin model to make it applicable for the dimer case. 
The three site term in Eq. \eqref{eq:fourspinham-trimer} obviously vanishes, while we extend the model with a six-spin term:  %$\vec{e}_i\vec{e}_j$. After these modifications the application of the multi-site Hamiltonian in \eqref{eq:4spin_multisite} for a dimer takes the form
\begin{equation}
    H=-J\vec{e}_1\vec{e}_2 +\dfrac{1}{2} B (\vec{e}_1\vec{e}_2)^2 + \dfrac{1}{3} C(\vec{e}_1\vec{e}_2)^3 \, ,
    \label{eq:fourspinham-dimer}
\end{equation}
where $J$ and $B$ are the previously introduced bilinear and biquadratic interactions, respectively, and $C$ is the so called bicubic interaction parameter. The two-spin variation of this Hamiltonian can easily be calculated to obtain the dimer analogue of Eq.~\eqref{eq:4spin-local} as
\begin{align}
    \mathcal{J}_{12}=&\left[-J+B(\vec{e}_1\vec{e_2})+ C (\vec{e}_1\vec{e_2})^2 \right] \mathcal{I}_3 \nonumber \\ &+ B (\vec{e}_2\circ\vec{e_1}) + 2C (\vec{e}_2\circ\vec{e_1})(\vec{e}_1\vec{e_2})\;.
    \label{eq:4spin-local-dimer}
\end{align}
From the interaction in Eq.~\eqref{eq:4spin-local-dimer} and the parametric configuration space path the local derivatives can be constructed and the same projection procedure can be applied as shown in Eq.~\eqref{eq:proj}. 

The perturbative method introduced in section \ref{sec:glob4spin} is, in principle, suitable to calculate the parameters $J$, $B$ and $C$, in particular, for the Fe dimer we found a very slow convergence for the multi-spin integrals. In this section, we therefore employ a different strategy to determine the global parameters by fitting the analytic form of the two-spin energy variation. The transverse derivatives along the path indicated in Fig.~\ref{fig:cu}(a) can be expressed as
\begin{align}
    \dfrac{\delta^2E_{12}}{\delta\vartheta_1\delta\vartheta_2} =& J - B - C -(2J-8B-14C)\sin^{2}{\left(\vartheta \right)} \, ,    \nonumber\\
    &-(8B+36C)\sin^{4}{\left(\vartheta \right)}+ 24 C \sin^{6}{\left(\vartheta \right)} \label{eq:dEthth-dimer}\\
    \dfrac{\delta^2E_{12}}{\delta\phi_1\delta\phi_2} =&\left( J- B \cos{\left(2 \vartheta \right)} -  C\cos^{2}{\left(2 \vartheta \right)} \right)\sin^2(\vartheta) \, \label{eq:dphph-dimer}\\
     \dfrac{\delta^2E_{12}}{\delta\vartheta_1\delta\phi_2} =&  \dfrac{\delta^2E_{12}}{\delta\phi_1\delta\vartheta_2} = 0 \;,
\end{align}
where the two non-zero curves can be used for the fit. To include information from both transverse derivatives, we perform simultaneous least square fits of the global parameters. The result of this procedure are collected in Tab.~\ref{tab:dimer-param}. Based on the estimated errors, the fit is more accurate for Cr$_2$/Cu(001), with an absolute error of approximately 0.1~meV. In contrast, the fit for Fe$_2$/Cu(001) exhibits larger uncertainties, with absolute errors of 4~meV and 6~meV for the parameters $B$ and $C$, respectively. In both cases we got a large bilinear term with opposite sign suggesting an anti-ferromagnetic ground state for the Fe dimer and a ferromagnetic one for the Cr dimer, in line with the observation in Ref. \cite{cardias2023unraveling}. The biquadratic interaction is by one magnitude smaller for both cases, but significantly larger for the Fe dimer. The bicubic interaction is relevant only for the Fe dimer. 

\begin{table}[htb!]
\caption{\label{tab:dimer-param}The global interaction parameters measured in meV for Cr and Fe dimers on Cu(001) obtained from fitting the two-spin variation of the energy. The numbers in the parentheses represent the absolute errors of the parameters.}
\begin{ruledtabular}
\begin{tabular}{c|rrr}
        & $J$ & $B$ & $C$ \\ \hline
     Fe$_2$/Cu(001) & $244(4)$ & $84(4)$& $-43(6)$ \\
     Cr$_2$/Cu(001) & $-283.1(0.1)$ &$35.5(0.1)$ &$0.9(0.2)$
\end{tabular}
\end{ruledtabular}
\end{table}

\begin{figure}[htb!]
    \centering
    \includegraphics[width=1.0\linewidth,trim= 0cm 0cm 0cm 0cm]{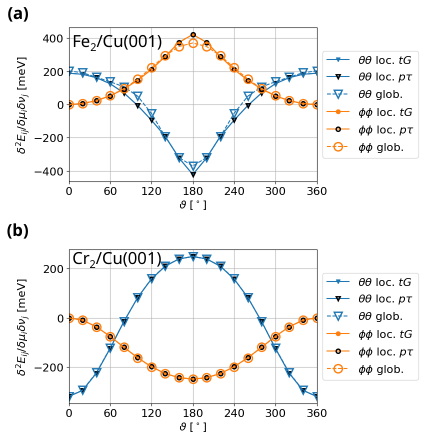}
    \caption{The transverse two-spin derivatives of the energy for \textbf{(a)} Fe$_2$/Cu(001)  and \textbf{(b} Cr$_2$/Cu(001) along the path sketched in Fig.~\ref{fig:cu}(a) as calculated using the two different local spin models and the fitted global spin model.
    }
    \label{fig:cu-der}
\end{figure}

Figure~\ref{fig:cu-der} shows the non-vanishing transverse derivatives obtained from the $tG$ and $p\tau$ formalisms, together with the corresponding analytic curves based on the fitted parameters in Tab.~\ref{tab:dimer-param}. Consistent with both the theoretical framework introduced above and the results for Cr$_3$/Au(111), the two local formalisms yield identical transverse derivatives. Moreover, the global model, Eq.~\eqref{eq:fourspinham-dimer}, reproduces the second derivatives remarkably well. Owing to the very small fitting errors in the multi-spin parameters, the agreement is almost perfect for the Cr dimer, as was also found for the Cr trimer on Au(111). For the Fe dimer, the larger discrepancy between the local and global models in the two-spin energy derivatives may indicate that higher-order terms omitted from the global model become significant.

\section{Discussion and conclusions}

%Summarizing the advances of this work, 
Based on non-relativistic density functional theory, we established the relationship between the local and the global spin models, resolving the ambiguities raised in the literature \cite{dos2021proper,cardias2022comment,dos2022reply}. In order to do so we started with the derivation of the global isotropic two- and four-spin interactions using a multiple scattering Green's function technique. 
%as introduced by Lászlóffy in Ref.~\cite{Laszloffy_dissertation} and we calculated the two-spin variation of this Hamiltonian. 
Then we derived the local two-spin variation of the energy within the same microscopic model in two ways, which we referred to as the $tG$ and the $p\tau$ formalism, and  
%following the calculations of Miklós Salánki in Ref.~\cite{salanki_bsc}, and we identified 
the local interaction parameters introduced by Cardias in Refs.~\cite{cardias2020dzyaloshinskii,cardias2020first}. 

In the collinear case the interpretation of the local derivatives is well established. For the relativistic case, Udvardi \textit{et al.} \cite{udvardi2003first} describe how one can construct the interaction matrix $\mathcal{J}_{ij}$ based on the second derivatives of the energy, requiring at least three linearly independent collinear configurations. In the non-relativistic case, this approach simplifies to a single calculation in any collinear state since the magnetic energy surface is isotropic and a unique Heisenberg-like interaction $J_{ij}$ is obtained \cite{Liechtenstein1987local}. On the contrary, by taking  non-collinear states as reference, even in the non-relativistic case non-Heisenberg-like interactions appear. These interactions are the DMI and the anisotropy in a non-collinear framework, see Eqs.~\eqref{eq:loc_dm} and \eqref{eq:loc_anis}.  Obviously, such interactions cannot happen in a global spin model and we refer to them as local interactions %since they describe energy variations in the vicinity of a given point in spin-configuration space. This is further reflected in the 
displaying a pronounced configuration dependence.

In Ref.~\cite{dos2021proper} it was already shown that the configuration dependence of the local interactions can be related to global four-spin interactions, however, this relationship was not directly shown in terms of calculating the four-spin and the local interactions on the same footing. In the present work, we showed results based on the implementation of the interaction parameters for an equilateral Cr trimer on Au(111) as a prototype system extended with the study of Fe and Cr dimers on Cu(001). The results for the local models
%of Miklós Salánki\cite{salanki_bsc} 
showed that the interaction parameters significantly depend on the formalism used for the calculation. 

The analysis of the transition between the two formalisms highlighted an important detail also emphasized by Udvardi \textit{et al.} \cite{udvardi2003first}, namely, that in a classical spin model the derivation of the interaction parameters should exclusively rely on transversal derivatives of the energy. By deriving local interactions from a global model we explicitly showed that longitudinal terms in the two-spin variation of the energy make the local interactions ambiguous.

In order to eliminate the longitudinal components from the interaction matrix we introduced projections to the transversal subspace in the local frame of reference. In terms of the global four-spin model containing three isotropic parameters ($J$, $B$ and $T$), the projection was performed analytically and we obtained local interaction parameters along a chosen path in the configuration space. These parameters showed an almost perfect agreement with numerical data derived from either of the $tG$ and $p\tau$ projected local spin models. Thus, we unambiguously demonstrated the direct relationship between the global and the local spin-model parameters. 

In Ref.~\cite{cardias2023unraveling}, Cardias {\em et al.} attempted to establish a quantitative relation between the (non-projected) local DMI and spin currents for magnetic dimers on non-magnetic substrates. They found that, to first order in the Dyson series, the time-reversal-invariant, spin-dependent part of the Green's function is proportional to $\vec{e}_1 \times \vec{e}_2$, which they associated with the spin current induced by the non-collinear spin structure. Since this contribution to the Green's function enters the expression for the local DMI, they concluded that the non-relativistic local DMI originates predominantly from spin currents.

Based on the global spin Hamiltonian given in Eq. (1) of the manuscript, the local DM interaction parameters can be expressed as
\begin{equation}
\vec{D}_{ij} = -2 \sum_{k,l} I_{ijkl}\left(\vec{e}_{k}\times\vec{e}_{l}\right) \, , \nonumber
\end{equation}
see also Eq.~(6) of Ref. [26]. This result can directly be compared to that of Ref.~\cite{cardias2023unraveling}. Importantly, however, the local DMI emerges naturally from the same theoretical framework used to derive the global spin Hamiltonian, without the need to invoke any additional microscopic interpretation of its origin.

We have to emphasize that a local spin model describes the transverse two-spin variations of the energy in the vicinity of a non-collinear ground state spin configuration.  
These energy variations can be fully described in terms of four transverse second derivatives of the energy. Reconstructing a $3 \times 3$ local interaction matrix from these four parameters does not provide any additional physical insight. Instead, the transverse energy derivatives can be directly employed to calculate the adiabatic spin-wave spectrum of non-collinear magnetic systems, as demonstrated for the 120$^\circ$ N\'eel structure of a Cr monolayer on Au(111) \cite{Balogh_2014}.

\section*{Data availability}
The data that support the findings of this study, including the raw values of global and local magnetic interaction parameters, are openly available at \cite{Nyari_2026_Zenodo}.

\begin{acknowledgments}
We acknowledge fruitful discussions with Stefan Blügel, Manuel dos Santos Dias, Samir Lounis, Attila Szilva, Levente Rózsa and László Udvardi.  This work was supported by the Ministry of Culture and Innovation and the National Research, Development and Innovation Office under No. K142652 and ADVANCED 149745. Bendeg\'uz Ny\'ari acknowledge funding by the Hungarian Academy of Sciences via the J\'{a}nos Bolyai Research Grant (Grant No. BO/00893/25). We acknowledge the Digital Government Development and Project Management Ltd. for awarding us access to the Komondor HPC facility based in Hungary.
\end{acknowledgments}

\appendix

\section{The evaluation of the two-spin and four-spin interactions}
\label{appendix1}
Here we give some more details of the derivation of the expressions \eqref{eq:Iij} and \eqref{eq:Iijkl}. By introducing the non-magnetic reference system of \eqref{eq:tspin}, \eqref{eq:lntau} can be split as
\begin{equation}
 \ln \boldsymbol{\tau}=\ln\mathbf{t}-\ln(\mathbf{I}-\mathbf{t}_r\mathbf{G}_0)-\ln(\mathbf{I}-\mathbf{t}_s\mathbf{G}_r) \, ,
\end{equation}
from which one can decompose the grand potential \eqref{eq:lloyd} as
\begin{equation}
    \Omega(\{ \vec{e}_i\} = \Omega_0 + \Omega_r + \sum_i \Omega_a (\vec{e}_i) + \Omega_\text{int}(\{ \vec{e}_i\}) \, .
\end{equation}
The first term is the constant term already present in \eqref{eq:lloyd}, the second one is a contribution of the non-magnetic reference system, the third one is an onsite anisotropy and the last one is the interaction term. Applying the Taylor series \eqref{eq:ln1mx}, the interaction term takes the following form,
\begin{equation}
    \Omega_\text{int}(\{ \vec{e}_i\} = -\dfrac{1}{\pi} \text{Im} \int_{-\infty}^{E_F} \text{d} \varepsilon \: \text{Tr} \sum_{k=1}^{\infty} \frac{1}{k} (\mathbf{t}_s \mathbf{G}_r)^k  \;.
\end{equation}
It can be shown that the terms corresponding to odd orders of $k$ vanish due to time reversal symmetry, so up to fourth order in $\mathbf{t}_s$ we have two terms only,
\begin{align}
    \Omega_\text{int}(\{ \vec{e}_i\}) \approx
    & -\dfrac{1}{\pi} \text{Im} \int_{-\infty}^{E_F} \text{d} \varepsilon \frac{1}{2}\text{Tr}\left[ \mathbf{t}_s \mathbf{G}_r\mathbf{t}_s \mathbf{G}_r\right] \nonumber \\
    &-\dfrac{1}{\pi} \text{Im} \int_{-\infty}^{E_F} \text{d} \varepsilon \frac{1}{4}\text{Tr}\left[ \mathbf{t}_s \mathbf{G}_r\mathbf{t}_s \mathbf{G}_r\mathbf{t}_s \mathbf{G}_r\mathbf{t}_s \mathbf{G}_r\right]\;.
\end{align}
Separating the trace over the spin and the orbital quantum number the two spin contribution is
\begin{align}
    \Omega_\text{int}^{(2)} =  & -\dfrac{1}{\pi} \text{Im} \int_{-\infty}^{E_F} \text{d} \varepsilon \frac{1}{2}\text{Tr}_L\left[ \mathbf{t}_s^{\alpha_1} \mathbf{G}_r\mathbf{t}_s^{\alpha_2} \mathbf{G}_r\right] \text{Tr}_S \left[ \sigma_{\alpha_1}\sigma_{\alpha_2}\right] \nonumber \\
    = & -\frac{1}{2}\sum_{i,j} \dfrac{2}{\pi} \text{Im} \int_{-\infty}^{E_F} \text{d} \varepsilon \text{Tr}_L\left[ \underline{t}_{s}^{i} \underline{G}_{r}^{ij}\underline{t}_s^{j} \underline{G}_{r}^{ij}\right] \vec{e}_i\vec{e}_j \, ,
    \label{eq:AA-Omega2}
%    \\ = &\frac{1}{2}\sum_{i,j} I_{ij}  \vec{e}_i\vec{e}_j\;,
\end{align}
where we used the identity, 
\begin{equation}
    \text{Tr}_S \left[ \sigma_{\alpha_1}\sigma_{\alpha_2}\right] = 2\delta_{\alpha_1\alpha_2} \, .
\end{equation} 
The definition of the two-spin integrals can easily be read off from \eqref{eq:AA-Omega2}, 
\begin{equation}
     I_{ij} = -\dfrac{2}{\pi} \text{Im} \int_{-\infty}^{E_F} \text{d} \varepsilon \text{Tr}_L\left[ \underline{t}_{s}^{i} \underline{G}_{r}^{ij}\underline{t}_s^{j} \underline{G}_{r}^{ij}\right]\;.
\end{equation}

Similar derivation applies for the four-spin interactions:
\begin{align}
    \Omega_\text{int}^{(4)} =  & -\dfrac{1}{\pi} \text{Im} \int_{-\infty}^{E_F} \text{d} \varepsilon \frac{1}{4}\text{Tr}_L\left[ \mathbf{t}_s^{\alpha_1} \mathbf{G}_r\mathbf{t}_s^{\alpha_2} \mathbf{G}_r\mathbf{t}_s^{\alpha_3} \mathbf{G}_r\mathbf{t}_s^{\alpha_4} \mathbf{G}_r\right] \nonumber \\ &  \qquad\qquad \text{Tr}_S \left[ \sigma_{\alpha_1}\sigma_{\alpha_2}\sigma_{\alpha_3}\sigma_{\alpha_4}\right] \nonumber \\
    = & - \frac{1}{4}\sum_{i,j,k.l} \dfrac{2}{\pi} \text{Im} \int_{-\infty}^{E_F} \text{d} \varepsilon \text{Tr}_L\left[ \underline{t}_{s}^{i} \underline{G}_{r}^{ij}\underline{t}_s^{j} \underline{G}_{r}^{jk}  \underline{t}_{s}^{k} \underline{G}_{r}^{kl}\underline{t}_s^{l} \underline{G}_{r}^{li} \right] \nonumber \\ & \qquad \quad [(\vec{e}_i\vec{e}_j) (\vec{e}_k\vec{e}_l) -  (\vec{e}_i\vec{e}_k) (\vec{e}_j\vec{e}_l) +  (\vec{e}_i\vec{e}_l) (\vec{e}_j\vec{e}_k)]  \, ,
     \label{eq:AA-Omega4}
%    \\ = & \frac{1}{4}\sum_{i,j,k.l} I_{ijkl}  [(\vec{e}_i\vec{e}_j) (\vec{e}_k\vec{e}_l) -  (\vec{e}_i\vec{e}_k) (\vec{e}_j\vec{e}_l) +  (\vec{e}_i\vec{e}_l) (\vec{e}_j\vec{e}_k)]\;,
\end{align}
where we used 
\begin{equation}
    \text{Tr}_S \left[ \sigma_{\alpha_1}\sigma_{\alpha_2}\sigma_{\alpha_3}\sigma_{\alpha_4}\right]  = \delta_{\alpha_1\alpha_2} \delta_{\alpha_3\alpha_4} - \delta_{\alpha_1\alpha_3} \delta_{\alpha_2\alpha_4} + \delta_{\alpha_1\alpha_4} \delta_{\alpha_2\alpha_3}
    \, .
\end{equation}
The four-spin interaction parameter is then given by
\begin{equation}
    I_{ijkl} = -\dfrac{2}{\pi} \text{Im} \int_{-\infty}^{E_F} \text{d} \varepsilon \text{Tr}_L\left[ \underline{t}_{s}^{i} \underline{G}_{r}^{ij}\underline{t}_s^{j} \underline{G}_{r}^{jk}  \underline{t}_{s}^{k} \underline{G}_{r}^{kl}\underline{t}_s^{l} \underline{G}_{r}^{li} \right] \, .
\end{equation}

\section{The $p\tau$ formalism of the local interactions}
\label{appendix:pTau}
Let us introduce the $p$-matrix as the inverse of the $t$-matrix,
\begin{equation}
    \underline{\underline{p}}^i = (\underline{\underline{t}}^i)^{-1} \, .
\end{equation}
The infinitesimal variation of the $t$-matrix can be expressed in terms of the infinitesimal variation of the $p$-matrix,
\begin{equation}
    \delta \underline{\underline{t}}^i = -\underline{\underline{t}}^i \delta \underline{\underline{p}}^i \underline{\underline{t}}^i \, .
    \label{eq:dt-dp}
\end{equation}
Within the non-relativistic theory, Eq.~\eqref{eq:dt-de} and the corresponding variation for the $p$ matrices apply. It can be shown that this is  compatible with Eq.~\eqref{eq:dt-dp} for infinitesimal rotations of the spin vectors $\vec{e}_i$. 

Substituting Eq.~\eqref{eq:dt-dp} into Eq.~\eqref{eq:dOmegaij2} and using the relationship between the SPO matrices and the structural Green's function matrices for $i \ne j$,
\begin{equation}
    \underline{\underline{\tau}}^{ij}= \underline{\underline{t}}^i \underline{\underline{G}}^{ij} \underline{\underline{t}}^j \, ,
\end{equation}
one can derive the equation for the two-spin variation of the grand potential in terms of the $p$- and the $\tau$-matrices,
\begin{equation}
\delta\Omega_{ij}= -\dfrac{1}{\pi} \Im \int_{-\infty}^{E_F} \dd \varepsilon \Tr ( 
          \delta \underline{\underline{p}}^i \underline{\underline{\tau}}^{ij}
          \delta \underline{\underline{p}}^j \underline{\underline{\tau}}^{ji} ) \, .
    \label{eq:dOmegaij-pt}
\end{equation}
This equation is formally identical to Eq.~\eqref{eq:dOmegaij2}, the $t$- and $G$-matrices being replaced by the $p$- and $\tau$-matrices, respectively. The interaction parameters can be derived in the same manner as before yielding
\begin{equation}
    \widehat{A}_{ij}^{\mu\nu}=-\dfrac{2}{\pi}\Im \int_{-\infty}^{E_F} \dd \varepsilon \Tr_L \left( 
    \underline{p}_s^i \underline{\tau}^{ij}_\mu
    \underline{p}_s^j \underline{\tau}^{ji}_\nu \right) \, ,
    \label{eq:Amat-pt}
\end{equation}
where $\mu,\nu \in \{0,x,y,z\}$ and
\begin{equation}
    \widehat{D}_{ij}^{\alpha}=-\dfrac{2}{\pi}\Re \int_{-\infty}^{E_F} \dd \varepsilon \Tr_L \left( \underline{\smash{p}}_s^i \underline{\tau}^{ij}_0  \underline{\smash{p}}_s^j \underline{\tau}^{ji}_\alpha-\underline{\smash{p}}_s^i \underline{\tau}^{ij}_\alpha  \underline{\smash{p}}_s^j \underline{\tau}^{ji}_0 \right) \; ,
    \label{eq:Dvec-pt}
\end{equation}
with $\alpha \in \{x,y,z\}$, being referenced as the $p\tau$ formalism of the local magnetic interactions.

\section{The ambiguity of the local interactions}
\label{appendix:locparamb}

The two-spin variation of the energy in the local spin models are separated into three terms \eqref{eq:loc_J}--\eqref{eq:loc_dm}. In \eqref{eq:local-deltaE_ij}, we deduced the local variation of the energy from the global spin model, but it requires further manipulation to make the mentioned separation visible. This can be achieved by adding terms to the variation of the energy proportional to $\vec{e}_i \delta\vec{e}_i$ and $\vec{e}_j \delta\vec{e}_j$, thereby leaving the variation unchanged. First we rewrite \eqref{eq:local-deltaE_ij} as
\begin{align}
\delta E_{ij}&=\left[ -J+B(\vec{e}_i\vec{e}_j)\right] (\delta\vec{e}_i\delta\vec{e}_j)\nonumber \\
&+\dfrac{1}{2} \left[ T\left( (\vec{e}_i\vec{e}_k)+ (\vec{e}_j\vec{e}_k)\right)\right]\delta\vec{e}_i \delta\vec{e}_j \nonumber \\
&+ B(\vec{e}_j\delta\vec{e}_i)(\vec{e}_i\delta\vec{e}_j) -  B(\vec{e}_i\delta\vec{e}_i)(\vec{e}_j\delta\vec{e}_j) \nonumber \\
&+\dfrac{1}{2} T \left(  (\vec{e}_k\delta\vec{e}_i)(\vec{e}_i\delta\vec{e}_j)+(\vec{e}_j\delta\vec{e}_i)(\vec{e}_k\delta\vec{e}_j)\right)  \nonumber\\
&-\dfrac{1}{2} T \left(  (\vec{e}_i\delta\vec{e}_i)(\vec{e}_k\delta\vec{e}_j)+(\vec{e}_k\delta\vec{e}_i)(\vec{e}_j\delta\vec{e}_j) \right)  \nonumber\\
&+\dfrac{1}{2} T (\vec{e}_k\delta\vec{e}_i)(\vec{e}_k\delta\vec{e}_j)  \, ,
\end{align}
which can further be rewritten as,
\begin{align}
\delta E_{ij}&=\left[ -J+B(\vec{e}_i\vec{e}_j)\right] (\delta\vec{e}_i\delta\vec{e}_j)\nonumber \\
&+\dfrac{1}{2} \left[ T\left( (\vec{e}_i\vec{e}_k)+ (\vec{e}_j\vec{e}_k)\right)\right]\delta\vec{e}_i \delta\vec{e}_j \nonumber \\
&+ B(\vec{e}_j\times\vec{e}_i) (\delta\vec{e}_i\times \delta\vec{e}_j) \nonumber \\
&+\dfrac{1}{2} T \left( (\vec{e}_k\times\vec{e}_i)+(\vec{e}_j\times\vec{e}_k) \right) (\delta\vec{e}_i\times \delta\vec{e}_j)  \nonumber\\
&+\dfrac{1}{2} T (\vec{e}_k\delta\vec{e}_i)(\vec{e}_k\delta\vec{e}_j)  \, ,
\label{eq:loc-dE_ij-v2}
\end{align}
where we used the vector identity $(\vec{a}\vec{b})(\vec{c}\vec{d})-(\vec{a}\vec{c})(\vec{b}\vec{d}) = (\vec{a}\times\vec{b})(\vec{c}\times\vec{d})$. From \eqref{eq:loc-dE_ij-v2} one can easily read off the local `isotropic', `anisotropy', and `DM' contributions to the energy variations. Substituting the parametric spin vectors \eqref{eq:local-spin-vectors} into \eqref{eq:loc-dE_ij-v2}, we obtain the local interaction parameters, 
\begin{equation}
J(\vartheta) = - J +  B + \dfrac{7}{6}T - \dfrac{3}{2} \left( B + T \right) \sin^2 \vartheta \, , 
\label{eq:jpar-glob-v2} 
\end{equation}
\begin{equation}
D_\perp(\vartheta) =  -\dfrac{\sqrt{3}}{2} (B - T) \sin^{2}\vartheta \, , 
\label{eq:dperp-glob-v2} 
\end{equation}
\begin{equation}
D_\parallel(\vartheta) =  - \dfrac{\sqrt{3}}{2} \left(B + \dfrac{T}{2}\right) \sin2\vartheta \, , 
\label{eq:dpar-glob-v2} 
\end{equation}
\begin{equation}
S_x (\vartheta) =  - \dfrac{1}{6}T \, , 
\label{eq:sx-glob-v2} 
\end{equation}
\begin{equation}
S_y (\vartheta) = - \dfrac{1}{6}T+\dfrac{1}{2}T\sin^{2}\vartheta \, , 
\label{eq:sy-glob-v2} 
\end{equation}
\begin{equation}
S_{yz} (\vartheta) = \dfrac{1}{4} T \sin2\vartheta \, .
\label{eq:syz-glob-v2}
\end{equation}
Note that the local DM interactions \eqref{eq:dperp-glob-v2}  and \eqref{eq:dpar-glob-v2}  are identical with those derived in Ref.~\cite{dos2021proper} and it has to be emphasized that the above expressions are different as compared to ~\eqref{eq:jpar-glob}--\eqref{eq:syz-glob}. Clearly enough, the reason for this is the inclusion of additional longitudinal terms in the energy variation. %With that we shade light to the fact that by including longitudinal terms, one can scale the value of the local spin model parameters to arbitrary values, while the energy variation remains the same.

%{\color{red} These results are almost identical with the one obtained by Lászlóffy, the only difference is that the trace of the $\vec{e}_k\circ\vec{e}_k$ term ($1/6T$) is not included in the isotropic term. }
\bibliography{references}% Produces the bibliography via BibTeX.

\end{document}